\documentclass[10pt]{iopart}
\usepackage[english]{babel}
\usepackage[T1]{fontenc}
\usepackage[utf8x]{inputenc}
\usepackage{iopams}
\usepackage{float}
\usepackage{amsmathe}
\usepackage[dvips]{graphicx}
\usepackage{auto-pst-pdf}
\usepackage{pstricks,pst-plot}
\usepackage{mathptmx}
\usepackage{helvet}
\usepackage{courier}

\newcommand{\E}{\rme}

\newcommand{\setintsect}{\operatorname{\cap}}
\newcommand{\setdiff}{\operatorname{\setminus}}
\newcommand{\setcard}{\operatorname{card}}
\newcommand{\set}[1]{\mathfrak{#1}}
\newcommand{\vect}[1]{\bi{#1}}
\newcommand{\matr}[1]{\mathbf{#1}}
\newcommand{\abs}[1]{\left\lvert#1\right\rvert}

\newcommand{\expect}{\operatorname{\mathsf{E}}}
\newcommand{\Var}{\operatorname{\mathsf{Var}}}
\newcommand{\Cov}{\operatorname{\mathsf{Cov}}}
\newcommand{\gl}[1]{\eref{eq-#1}}

\newcommand{\eqlbl}[1]{\label{eq-#1}}

\newcommand{\tab}[1]{\tref{tab-#1}}

\newcommand{\tbl}[1]{\label{tab-#1}}

\begin{document}

\title{A Bayesian approach to the linking of key comparisons}
\author{M Krystek}
\address{Physikalisch-Technische Bundesanstalt,\\ Bundesallee 100, D-38116 Braunschweig, Germany}
\eads{\mailto{Michael.Krystek@ptb.de}}
\author{H Bosse}
\address{Physikalisch-Technische Bundesanstalt,\\ Bundesallee 100, D-38116 Braunschweig, Germany}
\eads{\mailto{Harald.Bosse@ptb.de}}

\begin{abstract}
This contribution presents a Bayesian approach to the issue of linking of the results from key comparison measurements. A mathematical treatment based on Bayesian statistics for the analysis of the results from two comparisons with some joint participants is described. This robust statistical analysis provides expressions and standard uncertainties for the key comparison reference value (KCRV) and the degree of equivalence (DOE) as well as a conformity check without any assumption on a priority of one of the comparisons. In addition to the derivation of the mathematical formulae to be used for this type of `distributed linking', we also present one synthetic and one real linking example and discuss possible applications of this new linking procedure.
\end{abstract}


\section{Introduction}

The CIPM Mutual Recognition Arrangement (MRA) \cite{CIPM-MRA}, which was put into force in October 1999 and meanwhile was signed by 95 National Metrology Institutes (NMIs) and 4 international organisations worldwide describes the foundations and requirements for the mutual recognition of national measurement standards and for calibration and measurement certificates issued by NMIs. In addition the CIPM MRA covers a further 150 institutes designated by the signatory bodies. The MRA, chapter 3 states, that ``\emph{\dots the technical basis of this arrangement is the set of results obtained in the course of time through key comparisons carried out by the Consultative Committees of the CIPM, the BIPM, and the regional metrology organisations (RMOs), and published by the BIPM and maintained in the key comparison database. Key comparisons carried out by Consultative Committees or the BIPM are referred to as CIPM key comparisons; key comparisons carried out by regional metrology organisations are referred to as RMO key comparisons; RMO key comparisons must be linked to the corresponding CIPM key comparisons by means of joint participants. The degree of equivalence derived from an RMO key comparison has the same status as that derived from a CIPM key comparison.}'' To further support mutual confidence in the validity of calibration and measurement certificates issued by participating institutes so-called supplementary comparisons shall be carried out in addition. The metrology community has followed these documented requirements to carry out international comparison measurements \cite{CIPM-MRA-GUIDE} on a regular basis, whose results are continuously registered and made available to the public in the BIPM key comparison database (KCDB) \cite{KCDB}.

The issue of linking by means of joint participants is already addressed in the MRA, however without prescribing a particular mathematical linking procedure. Different approaches for robust statistical linking of the results of comparisons, e.~g. by transferring a key comparison reference value (KCRV) from one comparison to the other, have been discussed and proposed in literature and have been applied for the analysis of comparison results, too \cite{STEELE}, \cite{ELSTER}, \cite{WHITE}, \cite{SUTTON}.

Most of the proposed approaches for robust statistical linking follow the basic assumption of the MRA, namely that the results of an RMO key comparison have to be linked to the results of a CIPM key comparison. The CIPM key comparison in this concept generally is regarded as a primary comparison, both with respect to the chronological order and with respect to the  participating laboratories, because ``\emph{\dots participation in a CIPM key comparison is open to laboratories having the highest technical competence and experience, normally the member laboratories of the appropriate Consultative Committee.}'' (see sect. 6.1 of \cite{CIPM-MRA}). In this sense the linking of CIPM key comparisons and RMO key comparisons can be regarded as a hierarchical process. However, in \cite{STEELE} approaches for linking of comparisons through global minimisation by application of generalised least squares methods \cite{NIELSEN} were shortly discussed too, which can be regarded as examples of non-hierarchical linking methods.

In the field of dimensional metrology which is under the responsibility of the CIPM Consultative Committee for Length (CCL), international comparisons were started immediately after the MRA was signed. The results of the first CIPM key comparison on gauge blocks e.~g. were already published in 2002 \cite{THALMANN}.

Based on the experience gained from the international comparisons in the field of dimensional metrology, the CCL working group on dimensional metrology (CCL-WGDM\footnote{In June 2009 the working group structure of CCL has changed: MRA related work is now dealt with by the WG-MRA while linking issues are addressed by the task group on on KC linking (TG-L)}) has published a document which describes the characteristics of dimensional comparisons and which also proposes an alternative comparison scheme, the so-called CCL-RMO key comparison scheme \cite{CCL}. The background and the details of the proposed comparison scheme are not repeated here, however some core statements from this document are cited:
\begin{itemize}
  \item The CCL-RMO key comparisons follow the idea of several comparisons mutually linked together, without the necessity of a top level comparison delivering a KCRV.
  \item The classical scheme of CCL and RMO key comparisons can be considered to be a special case of the more general CCL-RMO scheme.
  \item In terms of linking laboratories within and across regions by calculation of their respective degrees of equivalence, the classical (hierarchical) and the CCL-RMO comparison scheme can be regarded to be equivalent. The linking of the comparisons is guaranteed by common participation of selected laboratories.
\end{itemize}

In \cite{DECKERJ} a generalised formalism was presented for linking a CIPM key comparison with similar regional key comparisons. As an example, this procedure was applied to link the results of a regional (SIM) gauge block comparison \cite{DECKER} to the results of the first CIPM key comparison on gauge blocks \cite{THALMANN}. In the applied procedure, a so-called linking invariant parameter is calculated based on the results of the linking laboratories, i.~e. those laboratories which participated in both comparisons, which in this case used different transfer standards.

This proposed method was discussed and well accepted in the CCL-WGDM, however it was also criticised that it still required to choose one comparison as primary. This condition is not in full alignment of the linking procedure to be applied for the analysis of comparisons according to the proposed CCL-RMO key comparison scheme. It was argued, that an adapted `distributed linking' procedure should be developed which should provide a statistically rigorous linking taking into account all the available information of the comparisons to be linked, however without the necessity to choose one comparison as a primary one.

In this paper a Bayesian approach for implementation of such a `distributed linking' is presented and an example for linking of two gauge block comparisons is discussed, which refers to the data published in \cite{THALMANN},\cite{DECKER}.

\section{The scenario}

Let us assume the following scenario: A group of laboratories (group A) has participated in a key comparison of a travelling standard A and another group of laboratories (group B) has participated in a key comparison of a travelling standard B. The task of each group was to measure one single measurand of the respective travelling standard (for example the length of a gauge block). In order to link the results of the two key comparisons, a certain group of laboratories (group C) has participated in both key comparisons, {i.\,e.} the laboratories of this group have been members of group A as well as of group B and thus have measured both travelling standards. All laboratories have reported the respective measurement results and the standard uncertainties associated with them. In addition, the laboratories of group C have reported the covariances associated with the measurement results of the two different travelling standards. It should further be assumed, that the standards have been stable, {i.\,e.} that they have not changed their properties in terms of time during the key comparisons.

In order to simplify the organisation of the data, we assume a labelling of the laboratories by assigning, independently of the membership in one of the groups, a unique natural number to each of them, starting by the number one in ascending order. In the following we will use the labels as indices for the respective data. It should be understood, although we are using an ordered set of indices, it is totally arbitrary which index is assigned to which laboratory. Any permutation of the indices used would also serve.

After having assigned the indices, we are able to refer to the index sets of the laboratories, rather than to the groups. Those laboratories, which have {e.\,g.} only measured the travelling standard A are within the index set $\set{I}_\text{A}\setdiff\set{I}_\text{B}$.

\section{The information available}

After having assigned labels (indices) to the laboratories, we are able to summarise the available data in a clear way just by using the same indices for the laboratories and the data. In the following, we will denote the measurement results concerning the travelling standard A obtained by the laboratories of group A by $x_{\text{A},i}$ $(i\in\set{I}_\text{A})$ and the standard uncertainties associated with this results by $u(x_{\text{A},i})$ $(i\in\set{I}_\text{A})$. Accordingly, we will denote the measurement results concerning the travelling standard B obtained by the  laboratories of group B by $x_{\text{B},i}$ $(i\in\set{I}_\text{B})$ and the standard uncertainties associated with this results by $u(x_{\text{B},i})$ $(i\in\set{I}_\text{B})$. It then only remains to deal with the covariances, associated with the measurement results concerning both travelling standards, as reported by the laboratories of the group C. In the following we will denote these covariances by $u(x_{\text{A},i},x_{\text{B},i})$ $\bigl(i\in(\set{I}_\text{A}\setintsect\set{I}_\text{B})\bigr)$. Equivalently we may use the correlation coefficients, defined by
\begin{equation}
r_{\text{A}\text{B},i}=\frac{u(x_{\text{A},i},x_{\text{B},i})}{u(x_{\text{A},i})u(x_{\text{B},i})},
\qquad i\in(\set{I}_\text{A}\setintsect\set{I}_\text{B}),
\eqlbl{1}
\end{equation}
instead of the covariances. We will frequently use the correlation coefficients as abbreviations in the subsequent calculations.

Under the prerequisite, that the two travelling standards under consideration have been stable during the key comparison, the assumption is justified, that all measurement results are the experimental realisation of only two different quantities, {i.\,e.} the measurands of the travelling standards A and B, denoted in the following by $Y_\text{A}$ and $Y_\text{B}$, respectively. In addition we assume, that the measurement results have already been corrected for possible systematic deviations in a suitable way and the uncertainties associated with the corrections have been included in the combined standard uncertainties associated with the corrected results.

Note, that we use upper case letters to denote the measurands and \emph{not} the respective measured value of a measurand, which we will denote by lower case letters throughout the text. We will however not differentiate between a measurand and its value (an ideal value to be estimated, sometimes called \emph{true value}) and use the same letter for both, since the respective meaning follows from the context.

\section{Establishing key comparison reference values}

\subsection{Preliminary remarks}

Using the ordinary arithmetic mean of the measured values to calculate a key comparison reference value (KCRV) can not be justified, because in doing so none of the known uncertainty values are taken into account and thus we do not make use of all available information, as required by the \emph{Guide to the Expression of Uncertainty in Measurement} (GUM) \cite{GUM}. Therefore we shall apply the Bayesian theory of measurement uncertainty \cite{WEISE-WOEGER}, which is entirely based on Bayesian statistics and the principle of maximum entropy (PME). The idea of this theory is first of all to establish a joint probability density function (pdf) of the measurands under consideration by taking into account the data obtained by measurement or otherwise given, together with their associated uncertainties, as well as physical relations and prior information about the measurands. This joint pdf expresses the state of incomplete knowledge about the measurands and is subsequently used to calculate the expectations of the measurands and its associated covariance matrix.

\subsection{The joint probability density function}

Since the measurements of all laboratories participating in the key comparison are assumed to be independent, we can assign probability density functions to the data of each laboratory individually. These pdfs belong to two different types only, because the data types of the groups A and B, excluding the data of group C, are similar and will thus lead to the same type of pdf. The data type of group C, however, is dissimilar from the data types of the groups A and B, excluding the data of group C, and hence we have to expect to get a different type of pdf for this group.

In order to assign pdfs\footnote{We will use here and throughout the document the abbreviation
$$
p(X|Y)=p_{X|Y}(x|y)
$$
for the conditional pdf, in order to simplify the notation of the formulae.} to the data, we follow a suggestion of Jaynes \cite{JAYNES1}, \cite{JAYNES2}, \cite{JAYNES3}, to use the principle of maximum entropy (PME) for this purpose. The PME requires to maximise the relative information entropy \cite{SHANNON}, \cite{KULLBACK} by a suitable pdf under the constraints imposed on the pdf by the normalisation condition and the data. In the process we make the usual assumption, that the measured data and the associated variances and covariances, respectively, may be equated with the respective expectations.

The PME yields under the circumstances at hand
\begin{equation}
p\bigl(x_{\text{A},i},u(x_{\text{A},i})\mid Y_\text{A}\bigr)=\frac{1}{u(x_{\text{A},i})\sqrt{2\pi}}
\exp\left(-\frac{(Y_\text{A}-x_{\text{A},i})^2}{2u^2(x_{\text{A},i})}\right)\,,
\qquad i\in\set{I}_\text{A}\setdiff\set{I}_\text{B}\,,
\eqlbl{4}
\end{equation}
\begin{equation}
p\bigl(x_{\text{B},i},u(x_{\text{B},i})\mid Y_\text{B}\bigr)=\frac{1}{u(x_{\text{B},i})\sqrt{2\pi}}
\exp\left(-\frac{(Y_\text{B}-x_{\text{B},i})^2}{2u^2(x_{\text{B},i})}\right)\,,
\qquad i\in\set{I}_\text{B}\setdiff\set{I}_\text{A}\,,
\eqlbl{5}
\end{equation}
and
\begin{multline}
p\bigl(x_{\text{A},i},x_{\text{B},i},u(x_{\text{A},i}),u(x_{\text{B},i}),r_{\text{A}\text{B},i}\mid Y_\text{A},Y_\text{B}\bigr)=
\frac{1}{2\pi\,u(x_{\text{A},i})u(x_{\text{B},i})\sqrt{1-r^2_{\text{A}\text{B},i}}}
 \\
\times\exp\left(-\frac{1}{2(1-r^2_{\text{A}\text{B},i})}
\left[\frac{(Y_\text{A}-x_{\text{A},i})^2}{u^2(x_{\text{A},i})}
-2\frac{r_{\text{A}\text{B},i}(Y_\text{A}-x_{\text{A},i})(Y_\text{B}-x_{\text{B},i})}
{u(x_{\text{A},i})u(x_{\text{B},i})}
+\frac{(Y_\text{B}-x_{\text{B},i})^2}{u^2(x_{\text{B},i})}\right]\right)\,, \\
\qquad i\in\set{I}_\text{A}\setintsect\set{I}_\text{B}\,.
\eqlbl{6}
\end{multline}
These pdfs can be esteemed as sampling distributions an thus be used to establish the likelihood function in the usual way. Note that the values $x_{\text{A},i}$ ($i\in\set{I}_\text{A}$) and $x_{\text{B},i}$ ($i\in\set{I}_\text{B}$) as well as their associated uncertainties $u(x_{\text{A},i})$ and $u(x_{\text{B},i})$, respectively, are given data, while $Y_\text{A}$ and $Y_\text{B}$ are unknown quantities.

The prior pdf $p(Y_\text{A},Y_\text{B})$ would allow us to express our knowledge about the quantities $Y_\text{A}$ and $Y_\text{B}$. However, in order to be conservative, we use Jeffreys' prior \cite{JEFFREYS} here, {i.\,e.} we assign a constant to this prior pdf. If more knowledge about  $Y_\text{A}$ and $Y_\text{B}$ would be available, the prior could be changed accordingly.

Using Bayes' theorem we obtain the posterior pdf
\begin{equation}
p(Y_\text{A},Y_\text{B}\mid\set{D})=C\E^{-\chi^2/2}\,,
\eqlbl{8}
\end{equation}
with the set $\set{D}$ representing all known data,
\begin{multline}
\chi^2=\sum_{i\in(\set{I}_\text{A}\setdiff\set{I}_\text{B})}
\frac{(Y_\text{A}-x_{\text{A},i})^2}{u^2(x_{\text{A},i})}
+\sum_{i\in(\set{I}_\text{B}\setdiff\set{I}_\text{A})}
\frac{(Y_\text{B}-x_{\text{B},i})^2}{u^2(x_{\text{B},i})}
\\
+\sum_{i\in(\set{I}_\text{A}\setintsect\set{I}_\text{B})}\frac{1}{1-r^2_{\text{A}\text{B},i}}
\left[\frac{(Y_\text{A}-x_{\text{A},i})^2}{u^2(x_{\text{A},i})}
-2\frac{r_{\text{A}\text{B},i}(Y_\text{A}-x_{\text{A},i})(Y_\text{B}-x_{\text{B},i})}
{u(x_{\text{A},i})u(x_{\text{B},i})}
+\frac{(Y_\text{B}-x_{\text{B},i})^2}{u^2(x_{\text{B},i})}\right]
\eqlbl{9}
\end{multline}
and a normalisation constant $C$. This is the posterior pdf of the quantities $Y_\text{A}$ and $Y_\text{B}$ given the data $\set{D}$.

After some algebraic transformations, equation \gl{9} can be written as
\begin{equation}
\chi^2=\frac{1}{1-\tilde{r}_{\text{A}\text{B}}^2}\left(
\frac{(Y_\text{A}-\hat{y}_{\text{A}})^2}{u^2(\hat{y}_{\text{A}})}
-2\tilde{r}_{\text{A}\text{B}}
\frac{(Y_\text{A}-\hat{y}_{\text{A}})(Y_\text{B}-\hat{y}_{\text{B}})}{u(\hat{y}_{\text{A}})u(\hat{y}_{\text{B}})}
+\frac{(Y_\text{B}-\hat{y}_{\text{B}})^2}{u^2(\hat{y}_{\text{B}})}\right)+q^2
\eqlbl{10}
\end{equation}
with
\begin{equation}
\hat{y}_{\text{A}}=\frac{bs_1+cs_2}{ab-c^2}
\eqlbl{11}
\end{equation}
\begin{equation}
\hat{y}_{\text{B}}=\frac{cs_1+as_2}{ab-c^2}
\eqlbl{12}
\end{equation}
\begin{equation}
u(\hat{y}_{\text{A}})=\sqrt{\frac{b}{ab-c^2}}
\eqlbl{13}
\end{equation}
\begin{equation}
u(\hat{y}_{\text{B}})=\sqrt{\frac{a}{ab-c^2}}
\eqlbl{14}
\end{equation}
\begin{equation}
\tilde{r}_{\text{A}\text{B}}=\frac{c}{\sqrt{ab}}
\eqlbl{15}
\end{equation}
\begin{multline}
q^2=\sum_{i\in(\set{I}_\text{A}\setdiff\set{I}_\text{B})}
\frac{(x_{\text{A},i}-\hat{y}_{\text{A}})^2}{u^2(x_{\text{A},i})}
+\sum_{i\in(\set{I}_\text{B}\setdiff\set{I}_\text{A})}
\frac{(x_{\text{B},i}-\hat{y}_{\text{B}})^2}{u^2(x_{\text{B},i})}
\\
+\sum_{i\in(\set{I}_\text{A}\setintsect\set{I}_\text{B})}\frac{1}{1-r^2_{\text{A}\text{B},i}}
\left(\frac{(x_{\text{A},i}-\hat{y}_{\text{A}})^2}{u^2(x_{\text{A},i})}
-2\frac{r_{\text{A}\text{B},i}(x_{\text{A},i}-\hat{y}_{\text{A}})(x_{\text{B},i}-\hat{y}_{\text{B}})}
{u(x_{\text{A},i})u(x_{\text{B},i})}
+\frac{(x_{\text{B},i}-\hat{y}_{\text{B}})^2}{u^2(x_{\text{B},i})}\right)
\eqlbl{16}
\end{multline}
and the abbreviations
\begin{equation}
a=\sum_{i\in(\set{I}_\text{A}\setdiff\set{I}_\text{B})}\frac{1}{u^2(x_{\text{A},i})}
+\sum_{i\in(\set{I}_\text{A}\setintsect\set{I}_\text{B})}
\frac{1}{(1-r^2_{\text{A}\text{B},i})u^2(x_{\text{A},i})}
\eqlbl{17}
\end{equation}
\begin{equation}
b=\sum_{i\in(\set{I}_\text{B}\setdiff\set{I}_\text{A})}\frac{1}{u^2(x_{\text{B},i})}
+\sum_{i\in(\set{I}_\text{A}\setintsect\set{I}_\text{B})}
\frac{1}{(1-r^2_{\text{A}\text{B},i})u^2(x_{\text{B},i})}
\eqlbl{18}
\end{equation}
\begin{equation}
c=\sum_{i\in(\set{I}_\text{A}\setintsect\set{I}_\text{B})}
\frac{r_{\text{A}\text{B},i}}{(1-r^2_{\text{A}\text{B},i})u(x_{\text{A},i})u(x_{\text{B},i})}
\eqlbl{19}
\end{equation}
\begin{equation}
s_1=\sum_{i\in(\set{I}_\text{A}\setdiff\set{I}_\text{B})}\frac{x_{\text{A},i}}{u^2(x_{\text{A},i})}
+\sum_{i\in(\set{I}_\text{A}\setintsect\set{I}_\text{B})}\frac{1}{1-r^2_{\text{A}\text{B},i}}
\left(\frac{x_{\text{A},i}}{u^2(x_{\text{A},i})}
-\frac{r_{\text{A}\text{B},i}x_{\text{B},i}}{u(x_{\text{A},i})u(x_{\text{B},i})}\right)
\eqlbl{20}
\end{equation}
\begin{equation}
s_2=\sum_{i\in(\set{I}_\text{B}\setdiff\set{I}_\text{A})}
\frac{x_{\text{B},i}}{u^2(x_{\text{B},i})}
+\sum_{i\in(\set{I}_\text{A}\setintsect\set{I}_\text{B})}\frac{1}{1-r^2_{\text{A}\text{B},i}}
\left(\frac{x_{\text{B},i}}{u^2(x_{\text{B},i})}
-\frac{r_{\text{A}\text{B},i}x_{\text{A},i}}{u(x_{\text{A},i})u(x_{\text{B},i})}\right)
\eqlbl{21}
\end{equation}
Note, that the auxiliary quantities $a$, $b$, $c$, $s_1$, $s_2$, and thereby the quantities $\hat{y}_{\text{A}}$, $\hat{y}_{\text{B}}$, $u(\hat{y}_{\text{A}})$, $u(\hat{y}_{\text{B}})$, $\tilde{r}_{\text{A}\text{B}}$ as well, depend only on the data obtained by the measurement. Thus $q^2$ does not depend on the quantities $Y_\text{A}$ and $Y_\text{B}$ and can hence be absorbed by the normalisation constant. Therefore, after renormalisation, we finally get
\begin{multline}
p(Y_\text{A},Y_\text{B}|\set{D})=
\frac{1}{2\pi u(\hat{y}_{\text{A}})u(\hat{y}_{\text{B}})\sqrt{1-\tilde{r}_{\text{A}\text{B}}^2}} \\
\times\exp\left[-\frac{1}{2(1-\tilde{r}_{\text{A}\text{B}}^2)}\left(
\frac{(Y_\text{A}-\hat{y}_{\text{A}})^2}{u^2(\hat{y}_{\text{A}})}
-2\tilde{r}_{\text{A}\text{B}}
\frac{(Y_\text{A}-\hat{y}_{\text{A}})(Y_\text{B}-\hat{y}_{\text{B}})}{u(\hat{y}_{\text{A}})u(\hat{y}_{\text{B}})}
+\frac{(Y_\text{B}-\hat{y}_{\text{B}})^2}{u^2(\hat{y}_{\text{B}})}\right)
\right]\,,
\eqlbl{22}
\end{multline}
{i.\,e.} the posterior pdf of the quantities $Y_\text{A}$ and $Y_\text{B}$ is a bivariate Gaussian pdf.

\subsection{The key comparison reference values}

Since the posterior pdf of the quantities $Y_\text{A}$ and $Y_\text{B}$ is a bivariate Gaussian pdf, it follows that
\begin{equation}
\expect\left[Y_\text{A}\right]=\hat{y}_{\text{A}}
\qquad\text\qquad
\expect\left[Y_\text{B}\right]=\hat{y}_{\text{B}}
\eqlbl{23}
\end{equation}
are the expectations of the quantities $Y_\text{A}$ and $Y_\text{B}$, respectively, and
\begin{equation}
\matr{U}_\vect{Y}=
\begin{pmatrix}
u^2(\hat{y}_{\text{A}}) & \tilde{r}_{\text{A}\text{B}}u(\hat{y}_{\text{A}})u(\hat{y}_{\text{B}}) \\
\tilde{r}_{\text{A}\text{B}}u(\hat{y}_{\text{A}})u(\hat{y}_{\text{B}}) & u^2(\hat{y}_{\text{B}})
\end{pmatrix}
\eqlbl{24}
\end{equation}
their associated covariance matrix. We now identify the values $\hat{y}_{\text{A}}$ and $\hat{y}_{\text{B}}$ as the key comparison reference values and the matrix $\matr{U}_\vect{Y}$ as their associated covariance matrix.

For convenience we summarise the results, using the covariances rather than the correlation coefficients.
\begin{equation}
a=\sum_{i\in(\set{I}_\text{A}\setdiff\set{I}_\text{B})}\frac{1}{u^2(x_{\text{A},i})}
+\sum_{i\in(\set{I}_\text{A}\setintsect\set{I}_\text{B})}
\frac{u^2(x_{\text{B},i})}{u^2(x_{\text{A},i})u^2(x_{\text{B},i})-u^2(x_{\text{A},i},x_{\text{B},i})}\,,
\eqlbl{25}
\end{equation}
\begin{equation}
b=\sum_{i\in(\set{I}_\text{B}\setdiff\set{I}_\text{A})}\frac{1}{u^2(x_{\text{B},i})}
+\sum_{i\in(\set{I}_\text{A}\setintsect\set{I}_\text{B})}
\frac{u^2(x_{\text{A},i})}{u^2(x_{\text{A},i})u^2(x_{\text{B},i})-u^2(x_{\text{A},i},x_{\text{B},i})}\,,
\eqlbl{26}
\end{equation}
\begin{equation}
c=\sum_{i\in(\set{I}_\text{A}\setintsect\set{I}_\text{B})}
\frac{u(x_{\text{A},i},x_{\text{B},i})}{u^2(x_{\text{A},i})u^2(x_{\text{B},i})-u^2(x_{\text{A},i},x_{\text{B},i})}\,,
\eqlbl{27}
\end{equation}
\begin{equation}
s_1=\sum_{i\in(\set{I}_\text{A}\setdiff\set{I}_\text{B})}\frac{x_{\text{A},i}}{u^2(x_{\text{A},i})}
+\sum_{i\in(\set{I}_\text{A}\setintsect\set{I}_\text{B})}
\frac{u^2(x_{\text{B},i})x_{\text{A},i}-u(x_{\text{A},i},x_{\text{B},i})x_{\text{B},i}}
{u^2(x_{\text{A},i})u^2(x_{\text{B},i})-u^2(x_{\text{A},i},x_{\text{B},i})}\,,
\eqlbl{28}
\end{equation}
\begin{equation}
s_2=\sum_{i\in(\set{I}_\text{B}\setdiff\set{I}_\text{A})}\frac{x_{\text{B},i}}{u^2(x_{\text{B},i})}
+\sum_{i\in(\set{I}_\text{A}\setintsect\set{I}_\text{B})}
\frac{u^2(x_{\text{A},i})x_{\text{B},i}-u(x_{\text{A},i},x_{\text{B},i})x_{\text{A},i}}
{u^2(x_{\text{A},i})u^2(x_{\text{B},i})-u^2(x_{\text{A},i},x_{\text{B},i})}\,.
\eqlbl{29}
\end{equation}
Key comparison reference values (KCRVs):
\begin{equation}
\hat{y}_{\text{A}}=\frac{bs_1+cs_2}{ab-c^2}\,,
\eqlbl{30}
\end{equation}
\begin{equation}
\hat{y}_{\text{B}}=\frac{cs_1+as_2}{ab-c^2}\,.
\eqlbl{31}
\end{equation}
Standard uncertainties of the KCRVs:
\begin{equation}
u(\hat{y}_{\text{A}})=\sqrt{\frac{b}{ab-c^2}}\,,
\eqlbl{32}
\end{equation}
\begin{equation}
u(\hat{y}_{\text{B}})=\sqrt{\frac{a}{ab-c^2}}\,.
\eqlbl{33}
\end{equation}
Covariance of the KCRVs:
\begin{equation}
u(\hat{y}_{\text{A}},\hat{y}_{\text{B}})=\frac{c}{ab-c^2}\,.
\eqlbl{34}
\end{equation}

\section{Degrees of equivalence}

The degrees of equivalence (DOE) are defined to be
\begin{equation}
d_{\text{A},i}=x_{\text{A},i}-\hat{y}_{\text{A}}\,,
\qquad i\in\set{I}_\text{A}\,,
\eqlbl{35}
\end{equation}
and
\begin{equation}
d_{\text{B},i}=x_{\text{B},i}-\hat{y}_{\text{B}}\,,
\qquad i\in\set{I}_\text{B}\,.
\eqlbl{36}
\end{equation}
These values give the deviations of each laboratory from the respective KCRV.

Following the rules for the propagation of variances, we obtain
\begin{equation}
u(d_{\text{A},i})=\sqrt{u^2(x_{\text{A},i})-u^2(\hat{y}_{\text{A}})}\,,
\qquad i\in\set{I}_\text{A}\,,
\eqlbl{37}
\end{equation}
and
\begin{equation}
u(d_{\text{B},i})=\sqrt{u^2(x_{\text{B},i})-u^2(\hat{y}_{\text{B}})}\,,
\qquad i\in\set{I}_\text{B}\,,
\eqlbl{38}
\end{equation}
for the respective uncertainties. The negative sign under the square root is due to the correlations $u(\hat{y}_{\text{A}},x_{\text{B},k})=u^2(\hat{y}_{\text{A}})$ and $u(\hat{y}_{\text{B}},x_{\text{B},k})=u^2(\hat{y}_{\text{B}})$ as shown in the appendix.

\section{Conformity test}

After the evaluation procedure has been accomplished, we have to check, whether the determined results do conform to the data given. This can be achieved in the following way. We observe, that on the one hand $\chi^2$ is represented by equation \gl{9} and on the other hand by equation \gl{10}. Thus, we can calculate the expectation $\expect\left[\chi^2\right]$ in two different ways, yielding
\begin{equation}
\expect\left[\chi^2\right]=N=2+q^2\,,
\eqlbl{42}
\end{equation}
with
\begin{equation}
N=\setcard\set{I}_\text{A}+\setcard\set{I}_\text{B}
\eqlbl{44}
\end{equation}
and $q^2$ given by equation \gl{16}. Thus, we obtain
\begin{equation}
q^2=N-2\,.
\eqlbl{43}
\end{equation}
Note that the right hand side of this equation may be regarded as the number of the degrees of freedom, because it is just the total number of measured data diminished by the number of estimated quantities.

The condition \gl{43} can, however, not be expected to be strictly fulfilled, but $N-2$ is the most probable value of $q^2$, because if the data are actually statistically distributed according to the model and the measurement uncertainties given, equation \gl{43} is the consequence. A significant deviation from this result is a signal that either the model is wrong or the data are suspect. Assuming the model to be correct, an outcome $q^2>N-2$ leads to the conclusion, that there is a strong possibility that either some of the measurement uncertainties have been underestimated or the respective measurement results contain uncorrected (or unknown) systematic deviations. Thus
\begin{equation}
q^2\le N-2
\eqlbl{45}
\end{equation}
can be taken as an indication of the conformity of the estimated results with the data.

\section{Examples}

In order to demonstrate the application of the method proposed in this paper, we show two examples. Our first example is a synthetic one. Data have been produced by a simulated sampling from Gaussian distributions as given in equations \gl{4} to \gl{6}, using the parameters\footnote{$\sigma$ and $\rho$ denote the standard deviation and the correlation coefficient, respectively, of a Gaussian distribution.} $Y_\text{A}=110$, $\sigma_\text{A}=20$, $Y_\text{B}=120$, $\sigma_\text{B}=50$, $\rho=0,5$ and a sample size of $n=50$. Subsequently the sampled data have been evaluated by using the usual formulae for the sample mean, the sample variance and the sample covariance. The resulting data thus obtained as well as their evaluation are shown in \tab{2a} and \tab{2b}.

\begin{table}[ht]
\caption{Data of the synthetic example obtained from the sampled data by application of the usual formulae for the sample mean, the sample variance and the sample covariance.}
\tbl{2a}
\begin{center}
\begin{tabular}{|l|rr|rr|r|}
\hline
Institute & $x_{\text{A},i}$ & $u(x_{\text{A},i})$ &
$x_{\text{B},i}$ & $u(x_{\text{B},i})$ & $r_{\text{A}\text{B},i}$ \\
\hline
LAB-01 & 	113,4 &  2,9 &  &  & \\
LAB-02 & 	112,1 &  2,8 &  &  & \\
LAB-03 & 	113,0 &  2,5 &  &  & \\
LAB-04 & 	110,6 &  2,6 &  &  & \\
LAB-05 & 	109,4 &  2,4 &  &  & \\
LAB-06 & 	107,0 &  2,6 &  &  & \\
LAB-07 & 	104,7 &  2,8 &  &  & \\
LAB-08 & 	109,0 &  2,6 &  &  & \\
\hline
LAB-09 & 	111,0 &  2,4 & 120,1 &  6,5 & 0,8 \\
LAB-10 & 	109,4 &  2,8 & 117,3 &  7,3 & 0,8 \\
LAB-11 & 	111,1 &  2,8 & 125,0 &  6,4 & 0,8 \\
LAB-12 & 	115,3 &  2,4 & 135,7 &  6,7 & 0,7 \\
\hline
LAB-13 & 	 & &129,7 &  6,1  & \\
LAB-14 & 	 & &129,1 &  7,5  & \\
LAB-15 & 	 & &125,0 &  7,1  & \\
LAB-16 & 	 & &123,6 &  6,6  & \\
LAB-17 & 	 & &123,0 &  6,9  & \\
\hline
\end{tabular}
\end{center}
\end{table}

\begin{table}[ht]
\caption{Degrees of equivalence for the deviation from the nominal values and their associated standard uncertainties, as well as the key comparison reference values and their associated standard uncertainties for the groups A and B, respectively, obtained by the application of the linking method proposed in this paper for the synthetic example.}
\tbl{2b}
\begin{center}
\begin{tabular}{|l|rr|rr|}
\hline
Institute & $d_{\text{A},i}$ & $u(d_{\text{A},i})$ & $d_{\text{B},i}$ & $u(d_{\text{B},i})$ \\
\hline
LAB-01 &  2,491 &  2,815 &  & \\
LAB-02 &  1,191 &  2,712 &  & \\
LAB-03 &  2,091 &  2,401 &  & \\
LAB-04 & -0,309 &  2,505 &  & \\
LAB-05 & -1,509 &  2,296 &  & \\
LAB-06 & -3,909 &  2,505 &  & \\
LAB-07 & -6,209 &  2,712 &  & \\
LAB-08 & -1,909 &  2,505 &  & \\
LAB-09 &  0,091 &  2,296 & -3,779 &  6,196 \\
LAB-10 & -1,509 &  2,712 & -6,579 &  7,030 \\
LAB-11 &  0,191 &  2,712 &  1,121 &  6,091 \\
LAB-12 &  4,391 &  2,296 & 11,821 &  6,405 \\
LAB-13 &  & &  5,821 &  5,775 \\
LAB-14 &  & &  5,221 &  7,238 \\
LAB-15 &  & &  1,121 &  6,822 \\
LAB-16 &  & & -0,279 &  6,300 \\
LAB-17 &  & & -0,879 &  6,614 \\
\hline
 & \multicolumn{4}{l|}{\rule{0pt}{4mm}$\hat{y}_{\text{A}}=110,909\,,\: u(\hat{y}_{\text{A}})=0,698$} \\
 & \multicolumn{4}{l|}{$\hat{y}_{\text{B}}=123,879\,,\: u(\hat{y}_{\text{B}})=1,966$} \\
 & \multicolumn{4}{l|}{\rule{0pt}{4mm}$q^2/(N-2)=0,89$} \\
\hline
\end{tabular}
\end{center}
\end{table}

As can be seen, the data do pass the conformity test. However, the KCRVs are slightly larger than the values used for the simulation. This is caused by the fact, that the correlation coefficients estimated from the sampled data tend to be too large compared to the correlation coefficient used for the simulation. This demonstrates the effect of small sample sizes.

As a second example data from the CCL-K1 \cite{THALMANN} and SIM.L-K1 \cite{DECKER} gauge block comparisons have been evaluated. The three institutes CENAM, NIST and NRC participated in both comparisons and hence serve as linking laboratories. For convenience, the results and their associated standard uncertainties, as reported in the respective publications, are repeated in \tab{2}. Values for the covariances have unfortunately not been reported, although systematic deviations present when measuring the gauge blocks in each of the linking laboratories with the same measuring system inevitably cause correlations, even if a correction for these systematic effects, as usually good practise, has been applied. Thus, following the recommendation given in the GUM in lack of this information\footnote{It should be noted, that this recommendation is consistent with the \emph{modus operandi} of the Bayesian theory, where only \emph{known} information is taken into account.}, all covariances have been assumed to be zero, knowing that this is certainly not true.

\begin{table}[ht]
\caption{Results for the deviations from the nominal length and their associated standard uncertainties as taken from the CCL-K1 \cite{THALMANN} and SIM.L-K1 \cite{DECKER} gauge block comparison reports (steel gauge block, nominal length 100 mm).}
\tbl{2}
\begin{center}
\begin{tabular}{|l|rr|rr|}
\hline
Institute & $x_{\text{A},i}/$nm & $u(x_{\text{A},i})/$nm & $x_{\text{B},i}/$nm & $u(x_{\text{B},i})/$nm \\
\hline
METAS &    -96,0 &   13,0 &  & \\
NPL &   -140,0 &   33,0 &  & \\
BNM-LNE &   -110,0 &   16,0 &  & \\
KRISS &   -104,3 &   20,6 &  & \\
NRLM &   -89,4 &   16,3 &  & \\
VNIIM &   -104,0 &   15,0 &  & \\
CSIRO &   -114,0 &   16,0 &  & \\
NIM &    -90,0 &   10,3 &  & \\
\hline
NIST &  -117,0 &   17,9 & -100,0 &   18,0  \\
CENAM &   -119,0 &   18,7 &  -93,0 &   23,0  \\
NRC &   -126,0 &   24,0 & -124,0 &   26,0  \\
\hline
INMETRO1 &  & & -98,0 &    \textbf{4,0} \\
INMETRO2 &  & & -68,0 &   29,0 \\
INTI &   & &-104,0 &   21,0 \\
CEM &    & &-148,0 &   17,0 \\
\hline
\end{tabular}
\end{center}
\end{table}

\begin{table}[ht]
\caption{Degrees of equivalence for the deviation from the nominal values and their associated standard uncertainties, as well as the key comparison reference values and their associated standard uncertainties for the groups A and B, respectively, obtained by the application of the linking method proposed in this paper (results obtained for the original data in \tab{2}).}
\tbl{3}
\begin{center}
\begin{tabular}{|l|rr|rr|}
\hline
Institute & $d_{\text{A},i}/$nm & $u(d_{\text{A},i})/$nm & $d_{\text{B},i}/$nm & $u(d_{\text{B},i})/$nm \\
\hline
METAS &    7,6 &   12,1 &  & \\
NPL &  -36,4 &   32,6 &  & \\
BNM-LNE &   -6,4 &   15,2 &  & \\
KRISS &   -0,7 &   20,0 &  & \\
NRLM &   14,2 &   15,6 &  & \\
VNIIM &   -0,4 &   14,2 &  & \\
CSIRO &  -10,4 &   15,2 &  & \\
NIM &   13,6 &    9,1 &  & \\
\hline
NIST &  -13,4 &   17,2 &    0,5 &   17,6 \\
CENAM &  -15,4 &   18,1 &   7,5 &   22,7 \\
NRC &  -22,4 &   23,5 &  -23,5 &   25,7 \\
\hline
INMETRO1 &  & &    2,5 &    1,7 \\
INMETRO2 &  & &   32,5 &   28,8 \\
INTI &  & &    -3,5 &   20,7 \\
CEM &  & &  -47,5 &   16,6 \\
\hline
 & \multicolumn{4}{l|}{\rule{0pt}{4mm}$\hat{y}_{\text{A}}=-103,6\,,\: u(\hat{y}_{\text{A}})=4,9$ nm} \\
 & \multicolumn{4}{l|}{$\hat{y}_{\text{B}}=-100,5\,,\: u(\hat{y}_{\text{B}})=3,6$ nm} \\
 & \multicolumn{4}{l|}{\rule{0pt}{4mm}$q^2/(N-2)=1,07$ (conformity test failed)} \\
\hline
\end{tabular}
\end{center}
\end{table}

\begin{table}[ht]
\caption{Degrees of equivalence for the deviation from the nominal values and their associated standard uncertainties, as well as the key comparison reference values and their associated standard uncertainties for the groups A and B, respectively, obtained by the application of the linking method proposed in this paper (results obtained after the uncertainty of the institute INMETRO1 has been increased from 4,0 nm to 11,2 nm).}
\tbl{4}
\begin{center}
\begin{tabular}{|l|rr|rr|}
\hline
Institute & $d_{\text{A},i}/$nm & $u(d_{\text{A},i})/$nm & $d_{\text{B},i}/$nm & $u(d_{\text{B},i})/$nm \\
\hline
METAS &    7,6 &   12,1 &  & \\
NPL &  -36,4 &   32,6 &  & \\
BNM-LNE &   -6,4 &   15,2 &  & \\
KRISS &   -0,7 &   20,0 &  & \\
NRLM &   14,2 &   15,6 &  & \\
VNIIM &   -0,4 &   14,2 &  & \\
CSIRO &  -10,4 &   15,2 &  & \\
NIM &   13,6 &    9,1 &  & \\
\hline
NIST &  -13,4 &   17,2 &    6,7 &   16,6 \\
CENAM &  -15,4 &   18,1 &   13,7 &   22,0 \\
NRC &  -22,4 &   23,5 &  -17,3 &   25,1 \\
\hline
INMETRO1 &  & &    8,7 &    8,9 \\
INMETRO2 &  & &   38,7 &   28,2 \\
INTI &  & &    2,7 &   19,9 \\
CEM &  & &  -41,3 &   15,6 \\
\hline
 & \multicolumn{4}{l|}{\rule{0pt}{4mm}$\hat{y}_{\text{A}}=-103,6\,,\: u(\hat{y}_{\text{A}})=4,9$ nm} \\
 & \multicolumn{4}{l|}{$\hat{y}_{\text{B}}=-106,7\,,\: u(\hat{y}_{\text{B}})=6,8$ nm} \\
 & \multicolumn{4}{l|}{\rule{0pt}{4mm}$q^2/(N-2)=1,00$} \\
\hline
\end{tabular}
\end{center}
\end{table}

The degrees of equivalence for the deviation from the nominal values and their associated standard uncertainties, as well as the key comparison reference values and their associated standard uncertainties for the groups A and B, respectively, obtained by the application of the linking method proposed in this paper are given in \tab{3} for the original data. It emerges, however, that the data do not pass the conformity test.

Assuming that the evaluation model is correct and that the respective measurement results have been corrected for systematic deviations, the test result leads to the conclusion, that there is a strong possibility that some of the measurement uncertainties have been underestimated. Inspecting the uncertainty values given in \tab{2}, it is obvious that ``\emph{\dots INMETRO1 submitted very optimistic uncertainty claims relative to conventional capabilities}'' \cite{DECKER} (the corresponding value is marked in bold face in the table). Since the gauge blocks have been measured at INMETRO by two completely different instruments, involving different staff members (data denoted by INMETRO1 represent results from a research grade instrument, while those denoted by INMETRO2 represent results from the routine gauge block calibration service offered by INMETRO) \cite{DECKER}, it is reasonable to question the very low uncertainty stated for INMETRO1. Therefore, this uncertainty value has been increased here from 4,0 nm to 11,2 nm, which is the smallest value in order that the data pass the conformity test. The respective results obtained with this change are given in \tab{4}.

A comparison of the results given in \tab{3} with those given in \tab{4} shows, that the reference value for group B and its associated standard uncertainty has noticeably changed. This strongly emphasises, that it is imperative to do a \emph{rigorous} conformity test. The $E_n$ criterion, for example, did not reveal any problem with the data from INMETRO1 during the SIM.l-K1 comparison. In fact the values $\abs{E_{95}}$ as well as $\abs{E}$ for the evaluation of the 100 mm steel gauge block data have even been comparable to those for the NRC \cite{DECKER}. The usual $\chi^2$ criterion at a 5 \% confidence level also fails to indicate a problem with the data.

\section{Conclusion}

In this contribution we have derived formulae for the linking parameters key comparison reference value (KCRV) and degree of equivalence (DOE) for analysis of two comparisons based on Bayesian statistics, {i.\,e.} taking all known information into account but without the necessity to choose one comparison as a primary one. The applicability of the linking approach was demonstrated on an example from gauge block metrology.

The proposed approach can be described as a type of `distributed linking', which has been looked for especially in the area of dimensional metrology within the working groups of the CCL.

This `distributed linking' method can, for example, be used to compare the results of two comparisons which were started in parallel using different transfer standards and where it does not seem meaningful to choose one comparison to be primary (defining a KCRV) and the other secondary (linking its comparison results to the KCRV of the first). Such comparisons with more than one loop have been organised if the number of interested participants became rather large, like e.~g. in the finished line scale comparison EUROMET.L-K7 with 31 participants \cite{EUROMET}.

In a strict Bayesian sense, using the `distributed linking' approach takes all known information from both comparisons into account to determine the linking parameters, i.~e. all the measurement results and their uncertainties. Instead, in application of the hierarchical linking approaches one could in principle argue that not all of the information is taken into account, because e.~g. the results gained from a RMO key comparison are not taken into account to reflect or re-analyse the results of a CIPM key comparison, although by the twofold involvement of the linking laboratories the level of knowledge is increased for both comparisons. On the other hand the results of a subsequent RMO comparison linked to a key comparison by the described `distributed linking' approach would also change the KCRV of the (already finished) key comparison. We see that this would impose new challenges for the operation of the KCDB.

The described approach can be extended to link more than two comparisons. Moreover, additional knowledge, as {e.\,g.} results from prior measurements, may be taken into account by choosing a suitable prior pdf.

\section{Acknowledgement}

The authors gratefully acknowledge the valuable comments of A. Balsamo from INRIM (Italy) and the suggestion to add the information given in appendix 2.

\section*{Appendix 1 (Calculation of covariances)}

In equations \gl{37} and \gl{38} the relations $u(\hat{y}_{\text{A}},x_{\text{B},k})=u^2(\hat{y}_{\text{A}})$ and $u(\hat{y}_{\text{B}},x_{\text{B},k})=u^2(\hat{y}_{\text{B}})$, respectively, have implicitly been used. In order to show that they are valid, we introduce the estimators
\begin{equation}
\hat{Y}_{\text{A}}=\frac{bS_1+cS_2}{ab-c^2}
\eqlbl{30b}
\end{equation}
and
\begin{equation}
\hat{Y}_{\text{B}}=\frac{cS_1+aS_2}{ab-c^2}\,,
\eqlbl{31b}
\end{equation}
where
\begin{equation}
S_1=\sum_{i\in(\set{I}_\text{A}\setdiff\set{I}_\text{B})}\frac{X_{\text{A},i}}{u^2(x_{\text{A},i})}
+\sum_{i\in(\set{I}_\text{A}\setintsect\set{I}_\text{B})}
\frac{u^2(x_{\text{B},i})X_{\text{A},i}-u(x_{\text{A},i},x_{\text{B},i})X_{\text{B},i}}
{u^2(x_{\text{A},i})u^2(x_{\text{B},i})-u^2(x_{\text{A},i},x_{\text{B},i})}
\eqlbl{28b}
\end{equation}
and
\begin{equation}
S_2=\sum_{i\in(\set{I}_\text{B}\setdiff\set{I}_\text{A})}\frac{X_{\text{B},i}}{u^2(x_{\text{B},i})}
+\sum_{i\in(\set{I}_\text{A}\setintsect\set{I}_\text{B})}
\frac{u^2(x_{\text{A},i})X_{\text{B},i}-u(x_{\text{A},i},x_{\text{B},i})X_{\text{A},i}}
{u^2(x_{\text{A},i})u^2(x_{\text{B},i})-u^2(x_{\text{A},i},X_{\text{B},i})}\,.
\eqlbl{29b}
\end{equation}
are functions of the random variables $X_{\text{A},i}$ ($i\in\set{I}_\text{A}$) and $X_{\text{B},i}$ ($i\in\set{I}_\text{B}$). We assume the measured values $x_{\text{A},i}$ ($i\in\set{I}_\text{A}$) and $x_{\text{B},i}$ ($i\in\set{I}_\text{B}$) to be a realisation of these random variables.

Since $\expect\left[X_{\text{A},i}\right]=Y_{\text{A}}$, $\expect\left[X_{\text{B},i}\right]=Y_{\text{B}}$, $\Var\left[X_{\text{A},i}\right]=u^2(x_{\text{A},i})$, $\Var\left[X_{\text{B},i}\right]=u^2(x_{\text{B},i})$ and $\Cov\left[X_{\text{A},i},X_{\text{B},i}\right]=u(x_{\text{A},i},x_{\text{B},i})$  ($i\in\set{I}_\text{A}$, $i\in\set{I}_\text{B}$)
it can be verified, that
\begin{equation}
	\expect\left[\hat{S}_1\right]=aY_{\text{A}}-cY_{\text{B}}\,,
	\qquad
	\expect\left[\hat{S}_2\right]=bY_{\text{B}}-cY_{\text{A}}\,,
\eqlbl{x1}
\end{equation}
as well as
\begin{equation}
	\Var\left[\hat{S}_1\right]=a\,,
	\qquad
	\Var\left[\hat{S}_2\right]=b\,,
	\qquad
	\Cov\left[\hat{S}_1,\hat{S}_2\right]=-c
\eqlbl{x2}
\end{equation}
is valid. Using these results, we obtain
\begin{equation}
	\expect\left[\hat{Y}_{\text{A}}\right]=\hat{y}_{\text{A}}\,,
	\qquad
	\expect\left[\hat{Y}_{\text{B}}\right]=\hat{y}_{\text{B}}\,,
\eqlbl{x3}
\end{equation}
\begin{equation}
	\Var\left[\hat{Y}_{\text{A}}\right]=u(\hat{y}_{\text{A}})\,,
	\qquad
	\Var\left[\hat{Y}_{\text{B}}\right]=u(\hat{y}_{\text{B}})\,,
	\qquad
	\Cov\left[\hat{Y}_{\text{A}},\hat{Y}_{\text{B}}\right]=u(\hat{y}_{\text{A}},\hat{y}_{\text{B}})\,,
\eqlbl{x4}
\end{equation}
{i.\,e.} the estimators $\hat{Y}_{\text{A}}$ and $\hat{Y}_{\text{B}}$ yield the correct results as given by equations \gl{30} to \gl{34}.

We observe that
\begin{equation}
	\Cov\left[\hat{S}_1,X_{\text{A},k}\right]=\Cov\left[\hat{S}_2,X_{\text{B},k}\right]=1
\eqlbl{x5}
\end{equation}
and
\begin{equation}
	\Cov\left[\hat{S}_1,X_{\text{B},k}\right]=\Cov\left[\hat{S}_2,X_{\text{A},k}\right]=0\,.
\eqlbl{x6}
\end{equation}
Thus
\begin{equation}
	u(\hat{y}_{\text{A}},x_{\text{A},k})=
	\Cov\left[\hat{Y}_{\text{A}},X_{\text{A},k}\right]=u^2(\hat{y}_{\text{A}})
\eqlbl{x7}
\end{equation}
and
\begin{equation}
	u(\hat{y}_{\text{B}},x_{\text{B},k})=
	\Cov\left[\hat{Y}_{\text{B}},X_{\text{B},k}\right]=u^2(\hat{y}_{\text{B}})\,.
\eqlbl{x8}
\end{equation}
This is what we wanted to show.

\section*{Appendix 2 (Uncorrelated measurement results)}

Here we consider the possibility that the measurement results of the linking laboratories (group C) have to be treated as if they were uncorrelated, because the respective covariances have not been reported. In this case we have to set all covariances $u(x_{\text{A},i},x_{\text{B},i})$ ($i\in(\set{I}_\text{A}\setintsect\set{I}_\text{B})$) equal to zero in the equations \gl{25} to \gl{29}. This yields
\begin{equation}
a=\sum_{i\in\set{I}_\text{A}}\frac{1}{u^2(x_{\text{A},i})}\,,
\eqlbl{x25}
\end{equation}
\begin{equation}
b=\sum_{i\in\set{I}_\text{B}}\frac{1}{u^2(x_{\text{B},i})}\,,
\eqlbl{x26}
\end{equation}
\begin{equation}
c=0\,,
\eqlbl{x27}
\end{equation}
\begin{equation}
s_1=\sum_{i\in\set{I}_\text{A}}\frac{x_{\text{A},i}}{u^2(x_{\text{A},i})}\,,
\eqlbl{x28}
\end{equation}
\begin{equation}
s_2=\sum_{i\in\set{I}_\text{B}}\frac{x_{\text{B},i}}{u^2(x_{\text{B},i})}\,.
\eqlbl{x29}
\end{equation}
Thus, we obtain from equations \gl{30} and \gl{31} for the key comparison reference values
\begin{equation}
\hat{y}_{\text{A}}=\frac{s_1}{a}
\qquad\text{and}\qquad
\hat{y}_{\text{B}}=\frac{s_2}{b}\,,
\eqlbl{x31}
\end{equation}
and from equations \gl{32} and \gl{33} for their associated standard uncertainties
\begin{equation}
u(\hat{y}_{\text{A}})=\frac{1}{\sqrt{a}}
\qquad\text{and}\qquad
u(\hat{y}_{\text{B}})=\frac{1}{\sqrt{b}}\,,
\eqlbl{x33}
\end{equation}
{i.\,e.} the usual weighted mean results of for the KCRVs separately for each group participating in the key comparisons. The results of the groups are independent of each other as indicated by the covariance of the KCRVs, $u(\hat{y}_{\text{A}},\hat{y}_{\text{B}})=0$, as obtained from equation \gl{34}. This demonstrates that the linking procedure essentially depends on the covariances reported by the linking laboratories.

\end{document}